\documentclass[10pt, conference]{IEEEtran}

\usepackage{enumerate}
\usepackage[utf8]{inputenc}
\usepackage{subcaption}
\usepackage{epsfig}
\usepackage{caption}
\usepackage{etoolbox}
\usepackage{wrapfig}
\usepackage{colortbl}
\usepackage{graphicx}

\usepackage{balance}
\usepackage{cite}
\usepackage[numbers,sort&compress]{natbib}
\usepackage{epsfig}
\usepackage{times}
\usepackage{graphicx}
\usepackage[figuresright]{rotating}
\usepackage{amssymb}
\usepackage{amsmath}
\usepackage{setspace}
\usepackage{rotating}
\usepackage{multirow}
\usepackage{wrapfig}
\usepackage{booktabs}
\usepackage{array}
\usepackage{comment}
\usepackage{pict2e}
\usepackage{epstopdf}
\usepackage{multirow}
\usepackage{hhline}
\usepackage{subcaption}
\usepackage[font=footnotesize]{caption} %
\usepackage[]{nohyperref}  %
\usepackage{url}  %

\usepackage{algorithmic}
\usepackage{algorithm}
\floatname{algorithm}{Algorithm} \textfloatsep 0.5cm

\usepackage{colortbl}

\newcommand{\be}{\begin{equation}}
\newcommand{\ee}{\end{equation}}

\usepackage[bottom]{footmisc}

\begin{document}
\title{Energy-efficient Wireless Analog Sensing for Persistent Underwater Environmental Monitoring}
\author{{\bf Vidyasagar Sadhu, Sanjana Devaraj, and Dario Pompili}\\
Department of Electrical and Computer Engineering, Rutgers University--New Brunswick, NJ, USA\\
E-mails: vidyasagar.sadhu@rutgers.edu, sd1049@scarletmail.rutgers.edu, pompili@ece.rutgers.edu\\
}

\maketitle

\thispagestyle{empty}
\pagestyle{plain}
\pagenumbering{gobble}

\begin{abstract}
The design of sensors or ``things'' as part of the new Internet of Underwater Things~(IoUTs) paradigm comes with multiple challenges including limited battery capacity, not polluting the water body, and the ability to track continuously phenomena with high temporal/spatial variability. We claim that traditional digital sensors are incapable to meet these demands because of their high power consumption, high complexity (cost), and the use of non-biodegradable materials. To address the above challenges, we propose a novel architecture consisting of a sensing substrate of dense analog biodegradable sensors over which lies the traditional Wireless Sensor Network~(WSN). The substrate analog biodegradable sensors perform Shannon mapping (a data-compression technique) using just a single Field Effect Transistor~(FET) without the need for power-hungry Analog-to-Digital Converters~(ADCs) resulting in much lower power consumption, complexity, and the ability to be powered using only sustainable energy-harvesting techniques. A novel and efficient decoding technique is also presented. Both encoding/decoding techniques have been verified via Spice and MATLAB simulations accounting for underwater acoustic channel variations. %
\end{abstract}

\section{Introduction}\label{sec:intro}
The Internet of Underwater Things~(IoUTs)~\cite{Domingo2012} is a novel class of IoTs enabling various practical applications in aqueous environments such as surveillance and real-time monitoring for pollution, environment and ocean current studies.

\textbf{Motivation:}
For the IoUTs to be a successful technology, the ``things'' or sensors should be able to capture temporal and spatial variations exhibited in various phenomena
in the underwater environment. Traditional digital sensors may not be the right candidates for sensing such phenomena, as they:
(i)~Have high power consumption, because of which they are put to sleep based on specific duty cycles; %
moreover, existing sensor encoding solutions use all-digital hardware, which demands high power and circuit complexity; as such, when the phenomenon exhibits high temporal variation, their batteries drain fast. %
(ii)~Are expensive, making them a costly choice for high-density deployment, which is needed to track phenomenon with high spatial variation. 
(iii)~May pollute when deployed in water bodies as the material used in the manufacturing of such sensors is not biodegradable; currently, most electronics are typically made with nondecomposable, nonbiocompatible, and sometimes even toxic materials, leading to serious ecological challenges. 

\begin{figure}
\begin{center}
\includegraphics[width=3.5in]{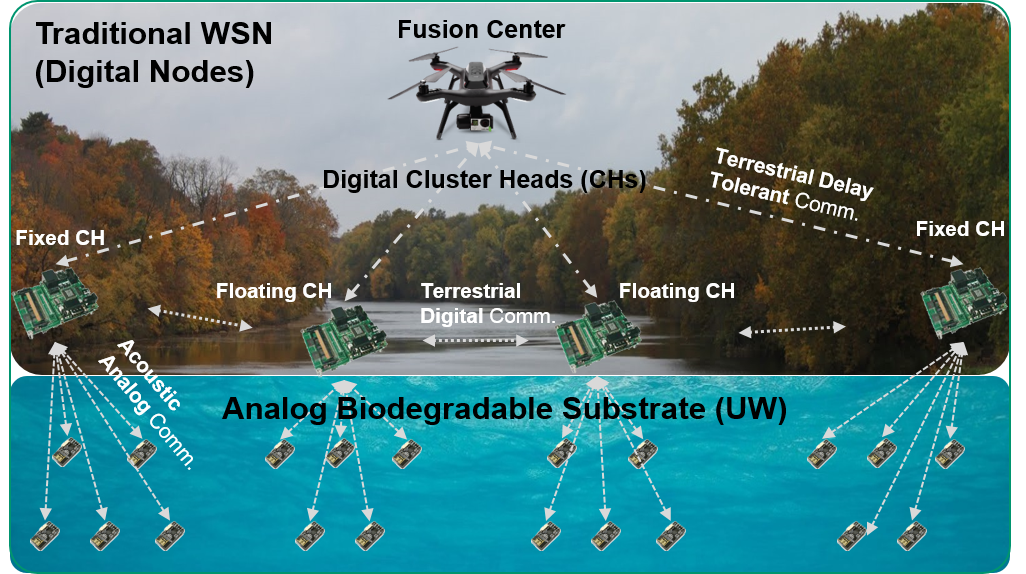}
\end{center}
\caption{A novel sensing architecture for real-time, persistent water monitoring using biodegradable analog sensors with AJSCC encoding capabilities as substrate above which lies the traditional Wireless Sensor Network~(WSN) consisting of digital Cluster Heads~(CHs) either drifting on the water or installed on the banks of the water body communicating among themselves and occasionally to a fusion center (e.g., drone).}\label{fig:UW_Architecture}
\vspace{-0.2in}
\end{figure}

\textbf{Our Vision:}
To track underwater phenomena with high temporal/spatial variation, there is a need to develop low-power and low-cost sensing devices~\cite{Zhao2018tbcas} that can be manufactured using biodegradable materials. To address the challenges above, we envision to design wireless transmitting sensors with Shannon-mapping~\cite{Shannon49} capabilities, a low-complexity technique for Analog Joint Source-Channel Coding~(AJSCC)~\cite{Hekland05}. We propose to develop ultra-low power realization of Shannon mapping using characteristics unique to semi-conducting biodegradable devices so that the sensors can be powered using only sustainable energy-harvesting techniques rather than polluting metal batteries. To realize this vision of dense IoUTs for persistent underwater monitoring, we propose a novel architecture shown in Fig.~\ref{fig:UW_Architecture}. The substrate consists of dense analog biodegradable sensors with Shannon-mapping capabilities and ultra-low power realizations using a \textit{single} Field Effect Transistor~(FET)-based encoding. These sensors transmit data continuously (no need to be put to sleep) to digital Cluster Heads~(CHs) in the traditional Wireless Sensor Network~(WSN), which decode the values received from analog sensors, i.e., they perform the reverse operation of Shannon mapping. These CHs also determine the optimum parameters of Shannon mapping %
by mining the received data (e.g., by estimating the temporal/spatial variation of the phenomenon being sensed by the substrate from the received values).

Moreover, since our design relies on energy-harvesting techniques to power the device, it cannot afford long-range transmission. This means that the sensors need to be deployed in high density and also close to CHs. This high-density deployment goes hand in hand with our low-complexity/low cost sensors where the encoding is performed using just a single transistor. As such, these CHs could be floating/drifting (on the water surface) along with the underwater analog sensors or be fixed on the banks of the water body (Fig.~\ref{fig:UW_Architecture}). Biodegradable MicroElectroMechanical Systems~(MEMS)-based acoustic transceivers with ranges of few meters are a perfect fit to our substrate sensors. %
Digital CHs send the processed information to a fusion center such as a drone as in a traditional WSN. Since the fusion center may not always be available, terrestrial delay-tolerant communication may need to be employed between CHs and center~\cite{Sadhu2017wons}.

\textbf{Related Work:}
All of the existing realizations of AJSCC are in the digital domain~\cite{Sadhu2017wons} except our previous work where we proposed an analog circuit realization of AJSCC~\cite{Zhao2018} using Voltage Controlled Voltage Sources~(VCVS) and Analog Dividers resulting in power consumption of the order of $90~\mu \rm{W}$. 
This is still high (considering additional overheads) to be powered with energy-harvesting techniques, which produce few $10$'s of $\mu \rm{W}$~\cite{IoUT}.
Toma et al.~\cite{EH_Piezo} proposed an underwater energy-harvesting system based on plucked-driven piezoelectric system with a
a maximum power density of $0.35~\mu W/{mm}^3$. Capitaine et al.~\cite{EH_MicrobialFC} proposed sedimentary microbial fuel cells as promising harvesting systems generating power of the order of few tens of $\mu W$, which is sufficient for powering our substrate sensors. Khan et al.~\cite{Khan2018} discuss energy-harvesting techniques that generate up to $100~\mu W$ from biochemistry and biochemical sources.

There is on going research work in the domain of biodegradable electronics~\cite{Irimia-Vladu2014}.
For example, totally disintegrable and biocompatible semiconducting polymers for thin-film transistors that are ultrathin ($<1 \mu \rm{m}$) and ultra lightweight ($\approx 2 g/m^2$) with low operating voltages are presented in~\cite{Lei5107}.
Bao's research group~\cite{Roberts2008} developed water stable Organic Field Effect Transistors~(OFETs) for aqueous sensing applications. 
There are also works in the domain of ultra-low power transceivers~\cite{Khan2018} and, in particular, biodegradable MEMS transceivers for our sensors. Kim et al.~\cite{Kim2015} and Rajavi et al.~\cite{Rajavi2016} present ultra-low power transistor-based transceiver designs with power consumption of the order of $50~\mu \rm{W}$ and a range of few meters. Amaya et al.~\cite{BioDegMEMS} present an efficient and environment friendly fabrication technology for MEMS transceivers using biodegradable polymer material applying hot embossing and polishing process. Mantha et al.~\cite{memstransceiver1} present a MEMS-based transceiver design that consumes about $50~\mu \rm{W}$. Power consumption of these transceivers can be supported by the energy-harvesting techniques mentioned above.
All these works suggest that our vision of using analog biodegradable sensors that are ultra lightweight, low power, and low cost in the sensing substrate is a feasible approach. %

\begin{figure}
\begin{center}
j\includegraphics[width=3.6in]{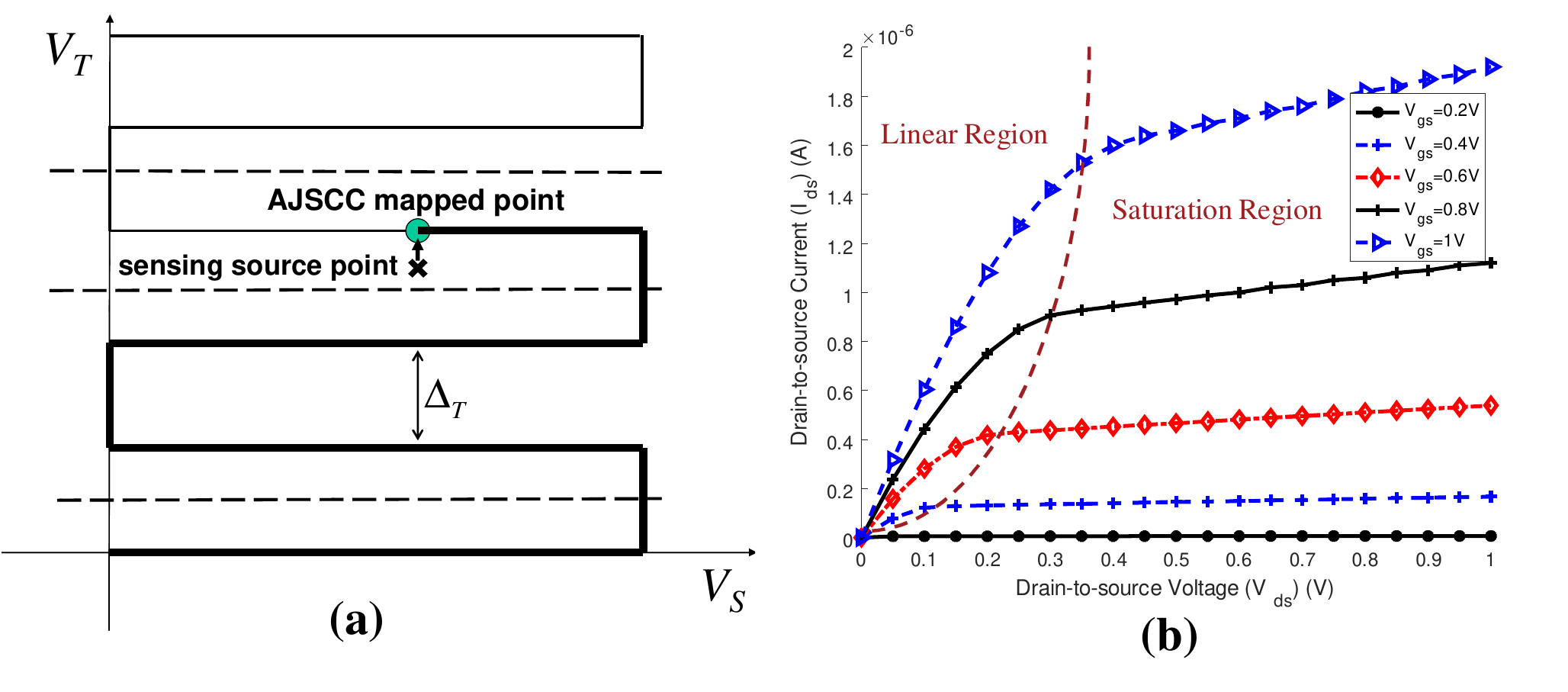}
\end{center}
\vspace{-0.2in}
\caption{(a)~Shannon rectangular mapping where the encoded value (of Turbidity $V_T$ and Salinity $V_S$ values) is the length of the curve from the origin to the mapped point. (b)~Shannon mapping realized via output characteristics ($I_{ds}$ vs. $V_{ds}$ for different $V_{gs}$) of a MOSFET in saturation region.
}\label{fig:ajscc_mos}
\vspace{-0.2in}
\end{figure}

\textbf{Our Contributions} can be summarized as follows:
\begin{itemize}
\item We propose the idea of using a FET's drain-source current ($I_{DS}$) vs. drain-source voltage ($V_{DS}$) for different values of gate-source voltage ($V_{GS}$) characteristic curves as a space-filling curve to perform Shannon mapping (2:1 compression), where $I_{DS}$ encodes $V_{DS}$ and $V_{GS}$ values.
\item We solve the challenge of non-unique mapping of multiple $V_{DS}$ and $V_{GS}$ values mapping to the same $I_{DS}$ value by developing a novel decoding technique at the receiver.
\item We validate the proposed encoding technique in terms of functionality, temporal and spatial variations of both phenomenon and channel, as well as average Mean Square Error~(MSE) using Spice and MATLAB simulations.
\end{itemize}

\textbf{Paper Outline:}
In Sect.~\ref{sec:prop_soln}, we present parts of our solution towards biodegradable analog sensing for persistent underwater environmental monitoring. In Sect.~\ref{sec:perf_eval}, we validate our design via both Spice and MATLAB simulations. Finally, in Sect.~\ref{sec:conc}, we conclude the paper and provide future directions.

\section{Proposed Solution}\label{sec:prop_soln}
We first introduce Analog Joint Source Channel Coding~(AJSCC); then, we preset our novel idea of using MOSFET to realize rectangular AJSCC, and discuss how to optimize the coding based on channel and phenomenon characteristics; finally, we describe how to solve the non-unique mapping of multiple pairs of $V_{gs},V_{ds}$ voltages to the same $I_{ds}$ value.

\textbf{Analog Joint Source Channel Coding~(AJSCC):}
AJSCC adopts Shannon mapping as its encoding method. Such mapping, in which the design of \emph{rectangular (parallel) lines} can be used for 2:1 compression
(Fig.~\ref{fig:ajscc_mos}(a)), was first introduced by Shannon in his seminal 1949 paper~\cite{Shannon49, Fresnedo13,Hekland05}. Later work has extended this mapping to a \emph{spiral type} as well as to N:1 mapping~\cite{Brante13}. AJSCC requires simple compression and coding, and low-complexity decoding. To compress the source signals (``sensing source point"), the point on the space-filling curve with minimum Euclidean distance from the source point is found (``AJSCC mapped point"), as in Fig.~\ref{fig:ajscc_mos}(a) where Salinity~(S) level is used for x-axis and Turbidity~(T) level for y-axis for illustration purposes. The two most-widely adopted mapping methods are rectangular and spiral shaped: in the former, the transmitted signal is the ``accumulated length'' of the lines from the origin to the mapped point; while in the latter it is the ``angle'' that \emph{uniquely} identifies the mapped point on the spiral. At the receiver (a digital CH), the reverse mapping is performed on the received signal using Maximum Likelihood~(ML) decoding. The error introduced by the two mappings is controlled by the spacing $\Delta_T$ between lines and spacing $\Delta_S$ ($S$ for spiral) between spiral arms, respectively: with smaller $\Delta_T$ (or $\Delta_S$), approximation noise is reduced; however, channel noise is increased as a little variation can push the received symbol to the next parallel line (or spiral arm) (i.e., large variation) instead of the little variation itself. 
In contrast, linear mapping techniques such as Quadrature Amplitude Modulation~(QAM) have errors spread on the constellation plane. Therefore channel noise has less effect on the error performance for Shannon mapping than
them.

\textbf{FET-based Encoding at Transmitter:}
As mentioned in Sect.~\ref{sec:intro}, power is of utmost importance in biodegradable sensors as they are powered using energy-harvesting techniques providing power in the order of $\mu W$. Hence, there is a need to realize a very simple implementation of AJSCC for substrate sensors. To this end, we present the idea of using a FET device's output characteristics to perform the encoding. 

The relationship between $V_{gs}$, $V_{ds}$, and $I_{ds}$ for an \emph{ideal} MOSFET in the saturation region (Fig.~\ref{fig:ajscc_mos}(b)) is expressed as, $I_{ds}=\frac{1}{2} \cdot \frac{W}{L} \cdot {\mu}C_{ox} \cdot (V_{gs}-V_{th})^2$,
where $W,L~[\rm{m}]$ are width and length of the MOSFET channel, $\mu~[\rm{m^2/Vs}]$ is the electron mobility in the channel, and $C_{ox}~[\rm{F/m^2}]$ is the oxide capacitance per unit area. In reality, however, due to a phenomenon called \emph{channel length modulation}~(CLM)~\cite{MOSChaLenMod}, 
$I_{ds}$ keeps increasing at a very slow rate (governed by $V_{gs}$ and other parameters) in the saturation region due to a reduction of the effective channel length. 
Consequently, the saturation-region relationship for a MOSFET with channel length modulation is actually,
\begin{equation} \label{eq:ids_clm}
I_{ds}=\frac{1}{2} \cdot \frac{W}{L} \cdot {\mu}C_{ox} \cdot (V_{gs}-V_{th})^2 \cdot (1+{\lambda} V_{ds}),
\end{equation} where $\lambda~[\rm{V^{-1}}]$ is the channel length modulation parameter. Figure~\ref{fig:ajscc_mos}(b) shows these $I_{ds}$ curves in the saturation region to the right of dashed line, generated via Spice simulation where
$V_{gs}$ is varied in the discrete set, $0.2,0.4,...,1~\rm{V}$ ($28~\rm{nm}$ Silicon technology model MOSFET is used for illustration purpose).
We can notice that the slope of the current curves increases as $V_{gs}$ increases, which we leverage to perform the decoding at the receiver as explained below.

We propose the idea of using these saturation region characteristics of a MOSFET with channel length modulation to fill the space (instead of using rectangular parallel lines) and perform the encoding, where $I_{ds}$ encodes the values of $V_{gs}$ and $V_{ds}$ (as opposed to extracting the length of the curve from origin to the mapped point, as in Fig.~\ref{fig:ajscc_mos}(a)). It should be noted that the \textit{shape} of the output characteristics ($I_{ds}$ vs. $V_{ds}$ for different $V_{gs}$) of a biodegradable MOSFET is similar to that of regular Silicon MOSFET (as shown in~\cite{Lei5107}). Moreover, the current generated of the polymer MOSFET in~\cite{Lei5107} is of the order of few $\mu \rm{A}$; with a supply voltage of $1~\rm{V}$, this will result in few $\mu\rm{W}$ of power consumption, which can be supported by energy-harvesting techniques. 
Similarly to Fig.~\ref{fig:ajscc_mos}(a), it is necessary to have a discrete set of y-axis ($V_{gs}$) values, and the actual y-axis value is mapped to the nearest value from the set and applied to the MOSFET to generate the encoded current. 

Ideally, any new space-filling curves should preserve these properties: ($i$)~they should achieve better trade-off between channel noise/compression and approximation noise; ($ii$)~they should be realizable using \emph{all-analog} components; and ($iii$)~they should result in a \emph{unique} mapping (i.e., two or more sensor values should map to \textit{only one} AJSCC encoded value). While the proposed MOSFET-based space-filling technique satisfies ($i$) and ($ii$), it violates ($iii$) as a given $I_{ds}$ value
could, in theory, be generated from multiple pairs of $V_{gs}$ and $V_{ds}$ values. It may be possible that ($iii$) is satisfied in some specific scenarios when the range of $\Delta_{V_{gs}}$ and/or $V_{ds}$ is restricted; however, it is not true in the general case.
This is problematic as it is difficult to decode the correct $V_{gs}$ at the receiver. In order to address this challenge, we propose a decoding technique at the receiver based on the previously received $I_{ds}$ value along with a slope-matching technique.

\textbf{Coding Optimization based on Channel and Phenomenon Characteristics:}
We assume our substrate analog sensors also have receive capability to obtain simple configuration information such as $\Delta_T$ from their digital CH and optimize the coding (we are unable to include the design of the analog circuit that achieves MOSFET encoding with variable $\Delta_T$ due to space limitations). The receiver decides the optimum $\Delta_T$ based on a tradeoff among different quantities including spatial and temporal correlation of the phenomenon being sensed, of the underwater acoustic channel, power budget available at the sensors, and minimum Mean Square Error~(MSE) requirements of the application. 
A smaller $\Delta_T$ results in better MSE at the receiver as the quantization error is minimized; however, it also results in higher power consumption at the sensor (due to additional hardware) as higher resolution is sought.
As mentioned earlier, a smaller $\Delta_T$ is more susceptible to acoustic channel noise as the decoded point may lie on a different line and vice-versa. Let us quantify the spatial and temporal correlation of the phenomenon being sensed using two variables---$s_p$ and $t_p$, respectively---where the former indicates the range/radius within which the phenomenon has spatial correlation/similarity while the latter indicates the time interval during which the phenomenon has temporal correlation/similarity. We assume the receiver estimates these parameters (hence, is aware of them); studying such techniques is outside the scope of this short paper. These parameters can be leveraged at the receiver to increase $\Delta_T$ (i.e., increase the quantization error), which can be compensated by averaging ($\overline{MSE}$) within both time $t_p$ and space $s_p$ (as values do not change significantly within these time windows/space ranges),
and thereby save on power. The acoustic wireless channel also plays a role in this process. Let us quantify the spatial and temporal correlation of the channel as $s_c$ and $t_c$, respectively, similarly to $s_p$ and $t_p$ defined above. Now, if $t_c < t_p$, \emph{time diversity} can help improve the MSE at the receiver; similarly, if $s_c < s_p$, \emph{space diversity} can help. If time and/or space diversity exist, it is possible to have good average $\overline{MSE}$ at the receiver. However, when both are absent, i.e., $t_c > t_p$ \emph{and} $s_c > s_p$ (i.e., neither time nor space diversity can be exploited), the effective acoustic channel condition determines the MSE at the receiver. We would like to mention that our analog sensors would employ the Frequency Position Modulation and Multiplexing~(FPMM) proposed in one of our previous works~\cite{Zhao2017a} to communicate with the digital CHs. FPPM consumes low power (of the order of few $\mu W$ theoretically) due to a very low SNR operating region at the receiver (about $-40~\rm{dB}$).

\textbf{Decoding at Receiver:}
We assume that the discrete set of $V_{gs}$ values 
used at the transmitter for encoding is known at the receiver. This is a valid assumption as the receiver decides the optimum $\Delta_T$ to be used by the transmitter. The decoding process relies on the assumption that physical values do not change abruptly and hence two consecutive received $I_{ds}$ values at the receiver will lie on the same $I_{ds}$ curve (i.e., corresponding to a particular $V_{gs}$ value). The probability of them lying on different $I_{ds}$ curves (i.e., corresponding to different $V_{gs}$ values) is low as the two consecutive values---(i)~would have sensed similar values (i.e., the sampling rate at the sensor is more than $1/t_p$), would have experienced similar acoustic channel conditions (i.e, the sampling rate at the sensor is more than $1/t_c$) and hence would belong to the same $I_{ds}$ curve. The challenge then lies in identifying the correct $V_{gs}$ value out of the discrete set of $V_{gs}$ values used at the transmitter using these two consecutive $I_{ds}$ values. For this purpose, we make use of a \emph{slope-matching technique},
which involves matching slopes calculated using two different methods---i.e., using~\eqref{eq:ids_clm} ($slope_1$) as well as the two-point formula ($slope_2$). From~\eqref{eq:ids_clm}, the slope of the $I_{ds}$ curve (for a given $V_{gs}$) can be calculated as, $\frac{\partial I_{ds}}{\partial V_{ds}} = \lambda \cdot \frac{1}{2} \cdot \frac{W}{L} \cdot {\mu}C_{ox} \cdot (V_{gs}-V_{th})^2 \approx \lambda \cdot I_{ds}$. Since the two consecutively received $I_{ds}^{(1)}$ and $I_{ds}^{(2)}$ are likely to lie on the same curve, the slope of that curve can then be approximated as $slope_{1} = \lambda \cdot (I_{ds}^{(1)} + I_{ds}^{(2)})/2$. Conversely, the slope using the two-point formula ($slope_2$) is found as follows. Since the receiver knows the set of $V_{gs}$ values used at the transmitter, each of these $V_{gs}$ values is substituted in~\eqref{eq:ids_clm} for both $I_{ds}^{(1)}$ and $I_{ds}^{(2)}$ to calculate the corresponding $V_{ds}$ values. For example, assume a set of five $V_{gs}$ values, which will result in five $V_{ds}$ values for $I_{ds}^1$ and five for $I_{ds}^2$. Next, five slopes are found using $I_{ds}^{(1)},I_{ds}^{(2)}$ values and the above $V_{ds}$ values using the two-point formula, $\frac{y_{2}-y_{1}}{x_{2}-x_{1}}$, resulting in 5 $slope_2$ values. Finally, $slope_2$ that best matches with $slope_1$ (via absolute difference) is chosen and the corresponding $V_{gs}$ is selected as the correct value; the corresponding $V_{ds}^{(2)}$ and $V_{ds}^{(1)}$ are found by substituting the decoded $V_{gs}$ in~\eqref{eq:ids_clm} for $I_{ds}^{(1)}$ and $I_{ds}^{(2)}$.

\begin{figure}
\begin{center}
\includegraphics[width=3.5in,height=1.8in]{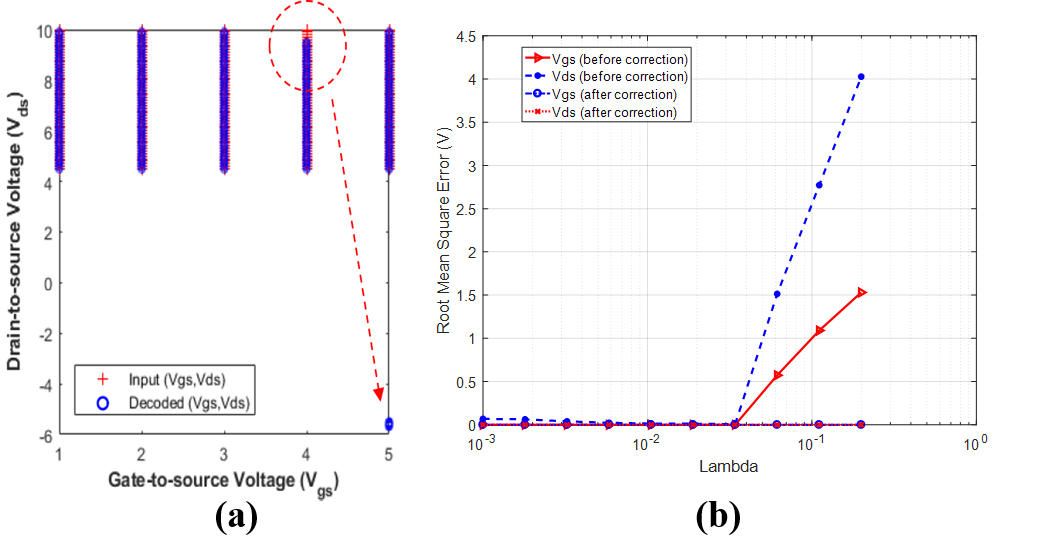}
\end{center}
\vspace{-0.2in}
\caption{(a)~Performance analysis graphs between input ($V_{gs}$, $V_{ds}$) and decoded ($V_{gs}$, $V_{ds}$) before correction; incorrect decodings have been resolved after correction (not shown). (b)~Error analysis graph of MSE of $V_{gs}$ and $V_{ds}$ for $\lambda$ ranging from $0.001$ to $0.2~\rm{V^{-1}}$ before and after correction.}\label{fig:mos_results}
\vspace{-0.2in}
\end{figure}

\section{Performance Evaluation}\label{sec:perf_eval}
To verify the functionality of our MOSFET-based encoding and its performance under different acoustic channel conditions, we have carried out Spice and MATLAB simulations.

\textbf{MOSFET Encoding and Decoding:}
To verify the MOSFET encoding and decoding functionality (without the effect of channel), we used a $0.18~\mu\rm{m}$ technology n-channel MOSFET~(nMOS) with $W\cdot\mu\cdot C_{ox}/L = 155\times 10^{-6}~\rm{F/Vs}$, $V_{th} = 0.74~\rm{V}$, $\lambda = 0.037~\rm{V^{-1}}$. $V_{ds}$ is varied from $5$ to $10~\rm{V}$ in increments of $0.1~\rm{V}$. The reason not to start from $0~\rm{V}$ is to ensure that the MOSFET is well into the saturation region. Five discrete values of $V_{gs}=1,2,3,4,5~\rm{V}$ are used; hence, for each $V_{gs}$, 50 values of $V_{ds}$ are considered. Upon applying these voltages to the MOSFET, the generated $I_{ds}$ values are recorded and sent to receiver (no channel) where the decoding process is done. At the receiver, each curve is processed independently and two consecutive $I_{ds}$ values from the same curve are used for decoding the correct $V_{gs}$ using the slope-matching technique. The results are shown in Fig.~\ref{fig:mos_results}(a), where the original values are shown using `+' and decoded values using `o'. We can see that most of the values are decoded correctly except a few ones indicated inside the circle that are decoded wrongly, as shown in the bottom-right corner. The reason is that the curves increase with a constant slope for a given $\lambda$, but a few current values are too closely spaced towards the end of the curve, which results in those values being mapped on to the immediate higher $V_{gs}$-curve. To solve this problem, we used a range-checking technique, where if the decoded $V_{ds}$ value corresponding to the best (in terms of slope match) $V_{gs}$ value does not fall within the $V_{ds}$ range assumed at the transmitter $(5,10)~\rm{V}$, the next best $V_{gs}$ value is chosen and the process is repeated iteratively. Using this correction, we are able to achieve $100\%$ accuracy in the decoding. 
We have observed that such incorrect decoding increases with $\lambda$ (CLM parameter), that can vary among different MOSFETs. To capture this effect, we varied $\lambda$ in the possible range $0.001$ to $0.2~\rm{V^{-1}}$ and plotted the MSE in Fig.~\ref{fig:mos_results}(b). We observe that for $\lambda > \approx 0.03$, the MSE increases rapidly before applying the correction and goes towards zero after applying correction. This ensures that our encoding/decoding technique is applicable to a wide range of MOSFETs, which can have different $\lambda$ values.

\begin{figure}
\begin{center}
\includegraphics[width=3.5in,height=1.8in]{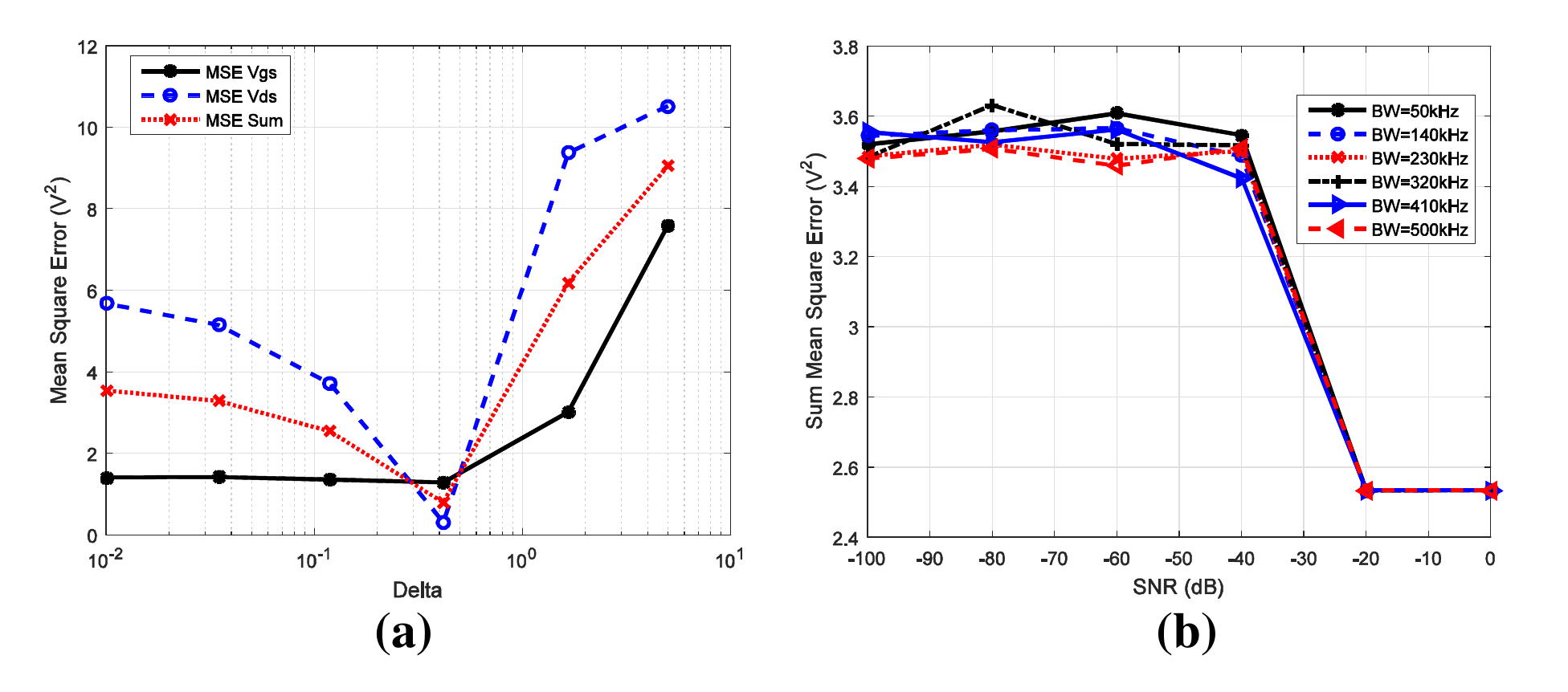}
\end{center}
\vspace{-0.2in}
\caption{(a)~Mean Square Error~(MSE) at receiver of $V_{gs}$, $V_{ds}$ and their sum vs. spacing between $I_{ds}$ curves ($\Delta$) at transmitter (Bandwidth = $410~\rm{kHz}$, SNR = $-20~\rm{dB}$). (b)~MSE vs. SNR for different bandwidths ($\Delta = 0.41$).}\label{fig:results_chan}
\vspace{-0.2in}
\end{figure}

\textbf{Variation with Channel Conditions:}
To quantify how MSE varies with channel conditions such as bandwidth~(BW) and SNR,
we ran MATLAB simulations with the following setting. We considered a $20 \times 20$ array of sensors generating values for 20 time instants (hence, %
$20 \times 20 \times 20$ values). We considered $s_p = t_p = 10$. This means that, at any given time instant, $10 \times 10$ subarrays (four in our case) have spatially similar values and, for each sensor, for a duration of $10$ time instants, the values are temporally similar. These temporally and spatially similar values are generated from random uniform distribution between $0$ and $1$. Two such instances have been considered, one for $V_{gs}$ and one for $V_{ds}$. Since the values generated are between 0 and 1, they are scaled and offset to lie between $(5,10)~\rm{V}$. This is done to capture the saturation behavior (in case of $V_{ds}$) and to ensure that $V_{gs}$ values are far from $V_{th}$. The error in decoding is very high when $V_{gs}$ is close to $V_{th}$ as the term $(V_{gs} - V_{th})^2$ appears in the denominator of $\partial V_{ds}/\partial V_{gs}$; hence, it is desirable to range $V_{gs}$ values far away from $V_{th}$. The $V_{gs}$ values have been quantized in accordance with Shannon mapping, before generating the encoded values using~\eqref{eq:ids_clm}. 
The encoded $I_{ds}$ values have been frequency modulated (i.e., each value is mapped to a frequency using, for example, a scaling factor) and then passed through a Rician channel in MATLAB with single path, and a Doppler shift equal to $2\%$ of the transmitting frequency. Then, Additive White Gaussian Noise~(AWGN) noise as per bandwidth~(BW) and SNR considered is added. At the receiver, the values are first passed through Fast Fourier Transform~(FFT) analysis to identify the peak frequency value, which is then mapped back to find the $\hat{I}_{ds}$ value using frequency demodulation (i.e., using the same scaling factor as in the transmitter). The slope-matching technique is used to decode the respective $\hat{V}_{gs}$ and $\hat{V}_{ds}$ values, and then $\overline{MSE}$ is found by averaging over both space ($s_p$) and time ($t_p$). Figure~\ref{fig:results_chan}(a) shows the $\overline{MSE}$ of $V_{gs}$, $V_{ds}$ and their sum for BW=$410~\rm{kHz}$ and SNR=$-20~\rm{dB}$. We can notice that indeed an optimum $\Delta^* = 0.41$ is achieved corresponding to $\overline{MSE}_{gs}=1.2~\rm{V^2}$ in $V_{gs}$, $\overline{MSE}_{ds} = 0.3~\rm{V^2}$ in $V_{ds}$ and $\overline{MSE}_{sum}=0.7~\rm{V^2}$ in their sum. Larger $\overline{MSE}_{ds}$ for smaller $\Delta$ is attributed to the decoding process. For a very small $\Delta$, it is possible that the decoded $\hat{V_{gs}}$ lies on adjacent levels to the actual one. However, since $\Delta$ is small, it will result in minor MSE for $V_{gs}$ but not for $\hat{V_{ds}}$ because of~\eqref{eq:ids_clm}. Figure~\ref{fig:results_chan}(b) shows the $\overline{MSE}_{sum}$ as the SNR is varied from $-100$ to $0~\rm{dB}$ for different bandwidths varying from $50$ to $500~\rm{kHz}$ with $\Delta^*=0.41$ found from above. We can notice that for $SNR<-30~\rm{dB}$ there is a sharp decrease in performance. While the performance is approximately similar for all bandwidths considered, we can notice an improvement in SNR as the bandwidth is increased.

\section{Conclusion and Future Work}\label{sec:conc}
We proposed a novel network architecture for the futuristic Internet of Underwater Things~(IoUTs) paradigm, where a simple Field Effect Transistor~(FET) based encoding is adopted to sense and transmit collected analog data, resulting in power efficient, low-complexity, and ultralight weight analog biodegradable transmitting sensors. Simulation results under different acoustic channel conditions indicate that such an encoding/decoding technique is highly promising for low bitrate IoUTs. Network simulations using ns-2, testing with realistic underwater channel models, and building a real prototype will be the focus of our future work.

\balance

\newpage
\bibliographystyle{IEEEtran} %
\bibliography{our_pubs,ref_ajscc_biodegradable}

\end{document}